\documentclass[aps,prc,showpacs,twocolumn,floatfix]{revtex4}

\usepackage{amsmath}
\usepackage{bm}
\usepackage{mathrsfs}
\usepackage{graphicx}
\usepackage{subfigure}

\renewcommand{\bar}[1]{\overline{#1}}
\newcommand{\ie}{{\it i.e.}}
\providecommand{\Journal}[4] {#1 {\bf#2}, #3 (#4)}
\providecommand{\PLB}{Phys. Lett. B} %
\providecommand{\PRD}{Phys. Rev. D}
\providecommand{\PRC}{Phys. Rev. C} %
\providecommand{\NPB}{Nucl. Phys. B} %
\providecommand{\EPJA}{Eur. Phys. J. A} %
\providecommand{\PR}{Phys. Rep.} %
\providecommand{\IJMPA}{Int. J. Mod. Phys. A } %

\begin{document}

\title{Analysis of the new Crystal Ball data on
$K^-p\to\pi^0\Lambda$
reaction with beam momenta of $514\sim  750$ MeV/$c$}
\author{Puze Gao, B.S.~Zou}
\affiliation{Institute of High Energy Physics, CAS, P.O. Box 918(4),
Beijing 100049, China} \affiliation{Theoretical Physics Center for
Science Facilities, CAS, Beijing 100049, China}
\author{A. Sibirtsev}
\affiliation{Helmholtz-Institut f\"{u}r Strahlen- und Kernphysik
(Theorie), Universit\"{a}t Bonn, D-53115 Bonn, Germany}
\affiliation{Jefferson Lab, 12000 Jefferson Avenue, Newport News,
Virginia 23606, USA}

\begin{abstract}
The Crystal Ball Collaboration has recently reported the
differential cross sections and $\Lambda$ polarization for the
reaction $K^-p\rightarrow \pi^0\Lambda$ using an incident $K^-$ beam
with momenta between 514 and 750 MeV/$c$. We make a partial wave
analysis for this process with an effective Lagrangian approach and
study the properties of some $\Sigma$ resonances around this energy
range. With the inclusion of the 4-star resonances $\Sigma(1189)$,
$\Sigma^*(1385)$, $\Sigma(1670){3\over 2}^-$, $\Sigma(1775){5\over
2}^-$, as well as a $\Sigma(1635){1\over 2}^+$, which is compatible
with the 3-star $\Sigma(1660){1\over 2}^+$ in PDG, our results can
well reproduce the experimental data. The parameters on the $\Sigma$
resonances and related couplings are studied.

\end{abstract}
\pacs{14.20.Jn, 25.20.Lj, 13.60.Le, 13.60.Rj}

\maketitle

\section{INTRODUCTION}

The $K^-p$ interactions at resonance region are important methods
for the study of resonance spectroscopy and interactions, especially
for hyperon with $S=-1$. Recently, the differential cross sections
as well as the $\Lambda$ polarization for
$K^-+p\rightarrow\pi^0+\Lambda$ are measured with very high
precision with the Crystal Ball spectrometer at the BNL Alternating
Gradient Synchrotron~\cite{prakhov09}, where neutron and photon
final states from $\pi^0\Lambda$ decays are well detected. The new
data provides a good opportunity for studying $\Sigma$-hyperon
resonances in the experimental energy range, which is between 514
and 750 MeV/$c$ for incident momentum, corresponding to
$\sqrt{s}=1569-1676$ MeV for c.m. energy.

The $\Sigma$-hyperon resonances in the Particle Data Group
(PDG)~\cite{PDG08} are mainly known from the analysis of $\bar K N$
reactions in the 1970s, and large uncertainties may exist not only
for the unestablished resonances with one or two stars, but also for
the established ones with three or four stars because of the limited
data and knowledge of background contributions. Moreover, there
still may be some new resonances that have not been discovered. Past
analyses of the reaction $\bar KN\rightarrow\pi\Lambda$ include the
energy dependent partial wave analysis with c.m. energy between 1540
and 2215 MeV~\cite{vanhorn75}, and the energy independent analysis
with c.m. energy between 1540 and 2150 MeV~\cite{baillon75}. Both
analyses considered the reaction amplitude parameterized as the sum
of resonance terms of Breit-Wigner form and a background term of
certain form. Different ways of background extraction may bring
large uncertainty to results.

In this work, benefitted from the available new data of high
precision, we make a partial wave analysis with an effective
Lagrangian approach. We aim at an improvement in the knowledge of
the $\Sigma$ resonances around the energy range concerned, as well
as their interactions with some other hadrons.

This paper is organized as follows. In section II, the theoretical
framework and amplitudes are presented for the reaction $\bar
KN\rightarrow\pi\Lambda$. In section III, the analysis results are
presented and compared with the experimental data, with some
discussions. In section IV, we give the summary and conclusion of
this work.

\section{THEORETICAL FRAMEWORK}

The effective Lagrangian method is an important theoretical approach
in describing various processes at resonance region, and is widely
used in partial wave analysis for the properties of resonances. For
the reaction $K^-+p\rightarrow\pi^0+\Lambda$, the Feynman diagrams
are shown in Fig.~1, where the incoming momenta are $k$ and $p$ for
kaon and proton, respectively, and the outgoing momenta are $q$ and
$p'$ for $\pi^0$ and the $\Lambda$, respectively. The main
contributions come from the t-channel $K^*$ meson exchange, the
u-channel proton exchange, and the s-channel $\Sigma$ and its
resonances exchanges. Note that in some previous analysis, the
t-channel and u-channel contributions were treated differently,
where they are treated as the background term with certain
parametrization.
\begin{figure}\label{fig:FMD2}
{\includegraphics[width=0.80\columnwidth]{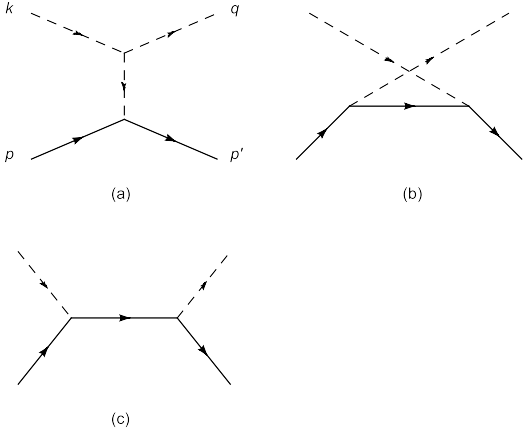}}
 \caption{Feynman diagrams for $K^-+p\rightarrow
\pi^0+\Lambda$. (a) t-channel $K^{*-}$ exchange; (b) u-channel
proton exchange; (c) s-channel $\Sigma$ resonances exchange.}
\end{figure}

For the t-channel $K^*$ meson exchange, the effective Lagrangian for
$K^*K\pi$ coupling is
\begin{equation}\label{kskpi}
{\cal L}_{K^*K\pi}=i g_{K^*K\pi} K^*_\mu({\pi\cdot\tau}\partial^\mu
K -\partial^\mu\pi\cdot\tau K)\,,
\end{equation}
where the isospin structure for $K^*K\pi$ is $\bar
K^*{\pi\cdot\tau}K$ with
\begin{equation}\label{isostr}
\bar K^*=(K^{*-},{\bar K}^{*0}), {\bf{\pi}}\cdot
{\bf\tau}=\left ( \begin{array}{cc} {\pi}^0& \sqrt{2}{\pi}^+ \\
\sqrt{2}{\pi}^- & -{\pi}^0
\end{array}\right ),K=\left(\begin{array}{c} K+
\\K^0\end{array}\right).
\end{equation}
Using the decay width $\Gamma_{K^*\rightarrow
K\pi}=50.8$~MeV~\cite{PDG08}, one gets the coupling constant
$g_{K^*K\pi}=-3.23$.

The effective Lagrangian for $K^*N\Lambda$ coupling is
\begin{equation}\label{ksnl}
{\cal L}_{K^*N\Lambda}=-g_{K^*N\Lambda}\bar\Lambda\big(\gamma_\mu
K^{*\mu}-{\kappa_{K^*N\Lambda}\over 2M_N}\sigma_{\mu\nu}\partial^\nu
K^{*\mu}\big)N+\mathrm{H.c.}\,,
\end{equation}
where $g_{K^*N\Lambda}$ and $\kappa_{K^*N\Lambda}$ are effective
coupling constants and can only be estimated from model predictions
or fit to some data. The popular potential model by Stoks and Rijken
gave two sets of these coupling constants~\cite{stoks99,oh06}:
\begin{eqnarray}
&&g_{K^*N\Lambda}=-4.26~~~~~\kappa_{K^*N\Lambda}=2.66~~~~\mathrm{(NSC97a)},\nonumber\\
&&g_{K^*N\Lambda}=-6.11~~~~~\kappa_{K^*N\Lambda}=2.43~~~~\mathrm{(NSC97f)}.
\end{eqnarray}
Thus we constrain $g_{K^*N\Lambda}$ between $-4.26$ and $-6.11$, and
$\kappa_{K^*N\Lambda}$ between 2.43 and 2.66 in our analysis. A
recent prediction from light cone QCD sum rules (LCSR) gives a
larger range for $g_{K^*N\Lambda}=-5.1\pm1.8$, while very different
values for $\kappa_{K^*N\Lambda}$~\cite{aliev09}. Some other works
for vector meson-baryon couplings also have large deviations on
$\kappa_{VBB}$~\cite{zhu99,erkol06,wang07}. For these uncertainties,
we also try the parameters in larger range and give some
discussions.

For the u-channel nucleon exchange, the effective Lagrangians are
\begin{equation}
{\cal L}_{\pi NN}={g_{\pi NN}\over 2M_N}{\bar
N}\gamma^\mu\gamma_5\partial_\mu\pi\cdot\tau N,
\end{equation}
\begin{equation}
{\cal L}_{KN\Lambda}={g_{KN\Lambda}\over
M_N+M_\Lambda}\bar{N}\gamma^\mu\gamma_5\Lambda\partial_\mu
K+\mathrm{H.c.},
\end{equation}
where $g_{\pi NN}=13.26$ and $g_{KN\Lambda}=-13.24$ are estimated
from flavor SU(3) symmetry relations~\cite{ohprt06,oh08}.

For the s-channel $\Sigma$ and its resonances exchange, we consider
effective couplings up to D-wave, which include intermediate states
with $J^P={1\over 2}^\pm$, ${3\over 2}^\pm$, and ${5\over 2}^-$.

For $\Sigma(1189)$ and its resonance with $J^P={1\over 2}^+$
contributions in s-channel , the effective Lagrangians are
\begin{equation}
{\cal L}_{KN\Sigma}={ g_{KN\Sigma}\over
M_N+M_\Sigma}\partial_\mu\bar{K}{\bf\bar\Sigma}\cdot\tau\gamma^\mu\gamma_5
N+\mathrm{H.c.},
\end{equation}
and
\begin{equation}
{\cal L}_{\Sigma\Lambda\pi}={g_{\Sigma\Lambda\pi}\over
M_\Lambda+M_\Sigma}\bar{\Lambda}\gamma^\mu\gamma_5\partial_\mu
\pi\cdot\Sigma+\mathrm{H.c.}.
\end{equation}
Where the isospin structure for $KN\Sigma$ coupling is
\begin{equation}\label{iso-kns}
\bar K=(K^-,{\bar K}^0), {\bf\bar{\Sigma}}\cdot
{\bf\tau}=\left ( \begin{array}{cc} {\bar\Sigma^0}& \sqrt{2}{\bar\Sigma^+} \\
\sqrt{2}{\bar\Sigma^-} & -{\bar\Sigma^0}
\end{array}\right ),N=\left(\begin{array}{c} p
\\n\end{array}\right).
\end{equation}
The coupling constants from SU(3) flavor symmetry relations predict
$g_{KN\Sigma}=3.58$ and $g_{\Sigma\Lambda\pi}=9.72$ for
$\Sigma(1189)$. With consideration of possible SU(3) symmetry
breaking effect, we multiply a tunable factor between $1/\sqrt{2}$
and $\sqrt{2}$ to the central value of
$g_{KN\Sigma}g_{\Sigma\Lambda\pi}$ in our analysis.

For intermediate $\Sigma$ state with $J^P={1\over 2}^-$, the
effective Lagrangians are
\begin{equation}
{\cal L}_{KN\Sigma({1\over 2}^-)}=-i g_{KN\Sigma({1\over
2}^-)}\bar{K}{\bf\bar\Sigma}\cdot\tau N+\mathrm{H.c.},
\end{equation}
and
\begin{equation}
{\cal L}_{\Lambda\pi\Sigma({1\over 2}^-)}=-i
g_{\Lambda\pi\Sigma({1\over
2}^-)}{\bar\Sigma}{\Lambda}\pi+\mathrm{H.c.}.
\end{equation}
The product of the coupling constants $g_{KN\Sigma({1\over
2}^-)}g_{\Lambda\pi\Sigma({1\over 2}^-)}$ is set to be a free
parameter in our analysis.

For intermediate $\Sigma^*$ state in s-channel with $J^P={3\over
2}^+$, the effective Lagrangians are
\begin{equation}\label{knss}
{\cal L}_{KN\Sigma^*}={f_{KN\Sigma^*}\over
m_K}\partial_\mu\bar{K}{\bf\bar{\Sigma}^{*\mu}}\cdot
{\bf\tau}N+\mathrm{H.c.}\,,
\end{equation}
and
\begin{equation}
{\cal L}_{\Sigma^*\Lambda\pi}={f_{\Sigma^*\Lambda\pi}\over
m_\pi}\partial_\mu\bar{\pi}\cdot{\bar{\Sigma}^{*\mu}}\Lambda+\mathrm{H.c.}\,,
\end{equation}
For $\Sigma^*(1385)$, the coupling constant
$f_{\Sigma^*\Lambda\pi}=1.27$ can be calculated from the decay width
$\Gamma_{\Sigma^*\rightarrow\Lambda\pi}\approx 31$ MeV~\cite{PDG08},
and $f_{KN\Sigma^*}=-3.22$ can be estimated from flavor SU(3)
symmetry relation~\cite{oh08}. With consideration of possible SU(3)
symmetry breaking effect, we multiply a factor between $\sqrt{2}$
and $1/\sqrt{2}$ as a free parameter to the central value of
$f_{KN\Sigma^*}$, and thus $f_{KN\Sigma^*}f_{\Sigma^*\Lambda\pi}$ is
constrained between -2.9 and -5.8 in our analysis.

For intermediate $\Sigma$ state in s-channel with $J^P={3\over
2}^-$, the effective Lagrangians are
\begin{equation}
{\cal L}_{KN\Sigma({3\over 2}^-)}={f_{KN\Sigma({3\over 2}^-)}\over
m_K}\partial_\mu\bar{K}{\bf\bar\Sigma}^\mu\cdot\tau \gamma_5
N+\mathrm{H.c.},
\end{equation}
and
\begin{equation}
{\cal L}_{\Lambda\pi\Sigma({3\over 2}^-)}={
f_{\Lambda\pi\Sigma({3\over 2}^-)}\over
m_\pi}\partial_\mu\pi{\bar\Sigma}^\mu\gamma_5{\Lambda}+\mathrm{H.c.}.
\end{equation}
The $\Sigma(1670)D_{13}$ is a four-star resonance in PDG. The above
coupling constants can be estimated from the decay width
$\Gamma_{\Sigma(1670)\rightarrow KN}$ and
$\Gamma_{\Sigma(1670)\rightarrow \pi\Lambda}$, which still have
large uncertainties. We constrain
$f_{KN\Sigma(1670)}f_{\Lambda\pi\Sigma(1670)}$ between $-1.5$ and
$-3.7$ in our analysis.

For intermediate $\Sigma$ state in s-channel with $J^P={5\over
2}^-$, the effective Lagrangians are
\begin{equation}
{\cal L}_{KN\Sigma({5\over 2}^-)}=g_{KN\Sigma({5\over
2}^-)}\partial_\mu\partial_\nu\bar{K}{\bf\bar\Sigma}^{\mu\nu}\cdot\tau
N+\mathrm{H.c.},
\end{equation}
and
\begin{equation}
{\cal L}_{\Lambda\pi\Sigma({5\over
2}^-)}=g_{\Lambda\pi\Sigma({5\over
2}^-)}\partial_\mu\partial_\nu\pi\cdot{\bar\Sigma}^{\mu\nu}{\Lambda}+\mathrm{H.c.}.
\end{equation}
The $\Sigma(1775)D_{15}$ is a four-star resonance in PDG. The
coupling constants can be estimated from the decay width
$\Gamma_{\Sigma(1775)\rightarrow KN}$ and
$\Gamma_{\Sigma(1775)\rightarrow \pi\Lambda}$. Because the mass of
$\Sigma(1775)$ is well above the range of the c.m. energy of the
experiments, the fit to the data is found insensitive to its
parameters, which are hence fixed to its central values in PDG. The
mass and width are fixed to be $1775$~MeV and 120~MeV, respectively,
and the product of the coupling constants
$g_{KN\Sigma(1775)}g_{\Lambda\pi\Sigma(1775)}$ is fixed to be
50~GeV$^{-4}$, corresponding to
$(\Gamma_{\pi\Lambda}\Gamma_{KN})^{1\over 2}/\Gamma_{\rm tot}\sim
-0.28$ in our analysis.

For each vertex of these channels, a form factor is attached to
describe the off-shell properties of the amplitudes.  For all the
channels considered, we adopt the form factor~\cite{oh08}
\begin{equation}\label{FB}
F_B(q_{ex}^2,M_{ex})={\Lambda^4\over
\Lambda^4+(q_{ex}^2-M_{ex}^2)^2}\, ,
\end{equation}
where the $q_{ex}$ and $M_{ex}$ are the 4-momenta and the mass of
the exchanged hadron, respectively.  The cutoff parameter $\Lambda$
is constrained between 0.8 and 1.5~GeV for all channels.

For the propagators with 4-momenta $p$, we use
\begin{equation}
{-g^{\mu\nu}+p^\mu p^\nu/ m_{K^*}^2\over p^2-m_{K^*}^2}
\end{equation}
for $K^*$ meson exchange ($\mu$ and $\nu$ are polarization index of
$K^*$);
\begin{equation}
{\not \! p}+m\over p^2-m^2
\end{equation}
for spin-1/2 propagator;
\begin{equation}
{{\not\! p}+m\over p^2-m^2}\Big(-g^{\mu\nu}+{
\gamma^\mu\gamma^\nu\over 3}+{\gamma^\mu p^\nu -\gamma^\nu
p^\mu\over 3m}+{2p^\mu p^\nu\over 3m^2}\Big)
\end{equation}
for spin-3/2 propagator; and
\begin{equation}
{{\not\! p}+m\over p^2-m^2}S_{\alpha\beta\mu\nu}(p,m)
\end{equation}
 for spin-5/2 propagator, where
\begin{eqnarray}
S_{\alpha\beta\mu\nu}(p,m)={1\over 2}({\bar g}_{\alpha\mu}{\bar
g}_{\beta\nu}+{\bar g}_{\alpha\nu}{\bar g}_{\beta\mu})-{1\over
5}{\bar g}_{\alpha\beta}{\bar g}_{\mu\nu}\nonumber\\
-{1\over 10}({\bar\gamma}_\alpha{\bar\gamma}_\mu{\bar
g}_{\beta\nu}+{\bar\gamma}_\alpha{\bar\gamma}_\nu{\bar
g}_{\beta\mu}+{\bar\gamma}_\beta{\bar\gamma}_\mu{\bar
g}_{\alpha\nu}+{\bar\gamma}_\beta{\bar\gamma}_\nu{\bar
g}_{\alpha\mu}),
\end{eqnarray}
with
\begin{eqnarray}
{\bar g}_{\mu\nu}=g_{\mu\nu}-{p_\mu p_\nu\over m^2},\nonumber\\
{\bar\gamma}_\mu=\gamma_\mu-{p_\mu\over m^2}{\not\! p}.
\end{eqnarray}
For unstable resonances, we replace the denominator $1\over p^2-m^2$
in the propagators by the Breit-Wigner form $1\over
p^2-m^2+im\Gamma$, and replace $m$ in the rest of the propagators by
$\sqrt{p^2}$. The $m$ and $\Gamma$ in the propagators represent the
mass and total width of a resonance, respectively.

The differential cross section for $K^-+p\rightarrow\pi^0+\Lambda$
at c.m. frame with $s=(p+k)^2$ can be expressed as
\begin{equation}
{d\sigma_{\pi^0\Lambda}\over d\Omega}={d\sigma_{\pi^0\Lambda}\over
2\pi d\cos\theta}={1\over 64\pi^2 s}{|{\bf q}|\over |{\bf
k}|}\bar{|{\cal M}|}^2,
\end{equation}
where $\theta$ denotes the angle of the outgoing $\pi^0$ relative to
beam direction in the c.m. frame, and $\bf k$ and $\bf q$ denote the
3-momenta of $K^-$ and $\pi^0$ in the c.m. frame, respectively. The
averaged amplitude square $\bar{|{\cal M}|}^2$ can be expressed as
\begin{eqnarray}
\bar{|{\cal M}|}^2&=&{1\over 2}\sum_{r_1,r_2}{|\cal M|}^2\nonumber\\
&=&{1\over 2}\mathrm{Tr}\big[({\not \! p'}+m_{\Lambda}){\cal
A}({\not \! p}+m_N)\gamma^0{\cal A^+}\gamma^0\big],
\end{eqnarray}
where $r_1$ and $r_2$ denote polarization of initial and final
state, respectively; and $p$ and $p'$ denote the 4-momenta of proton
and $\Lambda$ in the reaction. $\cal A$ is part of the total
amplitude, which can be expressed as
\begin{eqnarray}
{\cal M}=\bar u_{r_2}(p')~{\cal A}~u_{r_1}(p) =\bar
u_{r_2}(p')\big(\sum_i{\cal A}_i \big)u_{r_1}(p).
\end{eqnarray}
where $i$ denotes the $i$th channel that contributes to the total
amplitude.

The $\Lambda$ polarization in $K^-p\to\pi^0\Lambda\to\pi^0\pi N$ can
be expressed as
\begin{equation}
P_\Lambda={3\over \alpha_\Lambda} \Big (\int \cos\theta' {d\sigma_{
K^-p\to\pi^0\Lambda\to\pi^0\pi N}\over d\Omega d\Omega'}d\Omega'\Big
)\Big/ {d\sigma_{\pi^0\Lambda}\over d\Omega}
\end{equation}
where $\alpha_\Lambda=0.65$, and $d\Omega'=d\cos\theta'd\phi'$ is
the sphere space of the outgoing nucleon in the $\Lambda$ rest
frame, and $\theta'$ is the angle between the outgoing nucleon and
the vector $\bf{V}=\bf k\times\bf q$, which is vertical to the
reaction plane;

For $\Lambda\to \pi N$, the effective Lagrangian is
\begin{equation}
{\cal L}_{\Lambda\pi N}=G_F m_{\pi}^2{\bar N}(A-B\gamma_5)\Lambda,
\end{equation}
where $G_F$ denotes the Fermi coupling constant; $A$ and $B$ are
effective coupling constants, which can be calculated from the
parameters for $\Lambda$ decay ($\tau_\Lambda$, $\alpha_\Lambda$,
and $\Phi$) in PDG, and we take $A=1.762-0.238i$, $B=12.24$ in our
calculation.

The differential cross section for $K^-p\to\pi^0\Lambda\to\pi^0\pi
N$ can be expressed as
\begin{equation}
{d\sigma_{K^-p\to\pi^0\Lambda\to\pi^0\pi N}\over d\Omega
d\Omega'}={|{\bf q}||{\bf p}'_n||\bar{\cal{M}'}|^2\over 32^2 2\pi
m_\Lambda^2\Gamma_\Lambda s|\bf k|}
\end{equation}
where ${\bf p}'_n$ is the 3-momenta of the produced nucleon in the
$\Lambda$ rest frame, and $\Gamma_\Lambda=\tau^{-1}_\Lambda$ is
$\Lambda$ decay width; the amplitude $\cal M'$ is expressed as
\begin{equation}
{\cal M}'=\bar u_{r_3}(p_n)G_F m_\pi^2 (A-B\gamma_5)({\not \!
p'}+m_{\Lambda})\big(\sum_i{\cal A}_i \big)u_{r_1}(p)
\end{equation}
and $|\bar{\cal{M}'}|^2={1\over 2}\sum_{r_1,r_3}{\cal M'}{\cal
M'^+}$.

\section{RESULTS AND DISCUSSIONS}

In this analysis, the t-channel $K^*$ exchange and the u-channel
proton exchange amplitudes are fundamental ingredients, which are
different from previous analyses using some polynomial
parametrization for background contributions and are more physically
based, although there are also some parameters to be fitted in
reasonable ranges. The $\Sigma(1189){1\over 2}^+$,
$\Sigma^*(1385){3\over 2}^+$, $\Sigma(1670){3\over 2}^-$ and
$\Sigma(1775){5\over 2}^-$ contributions in s-channel are always
included in our analysis, partly because these channels should
contribute to the reaction by the knowledge of their existence as
well established four-star resonances, partly because the present
data favor the inclusion of them. Still some parameters in the above
channels have uncertainties and are to be fitted in the analysis.
The ranges of the parameters have been constrained from the PDG
estimates or model predictions, which have been explained in section
II. From the above 6 channels of 14 tunable parameters constrained
in the allowed range, the best fit to the differential cross
sections and the $\Lambda$ polarization gives a $\chi^2$ of about
763 for the total 248 data points. The results are shown by the
(blue) dashed lines in Fig.~2 and Fig.~3 for the differential cross
sections and the $\Lambda$ polarization, respectively. Although the
fit looks already quite good qualitatively, from detailed comparison
with the very precise data and the quite large $\chi^2$, some
systematic deviations still exist.

\begin{figure*}\label{fig:DCS}
{\includegraphics[width=1.3\columnwidth]{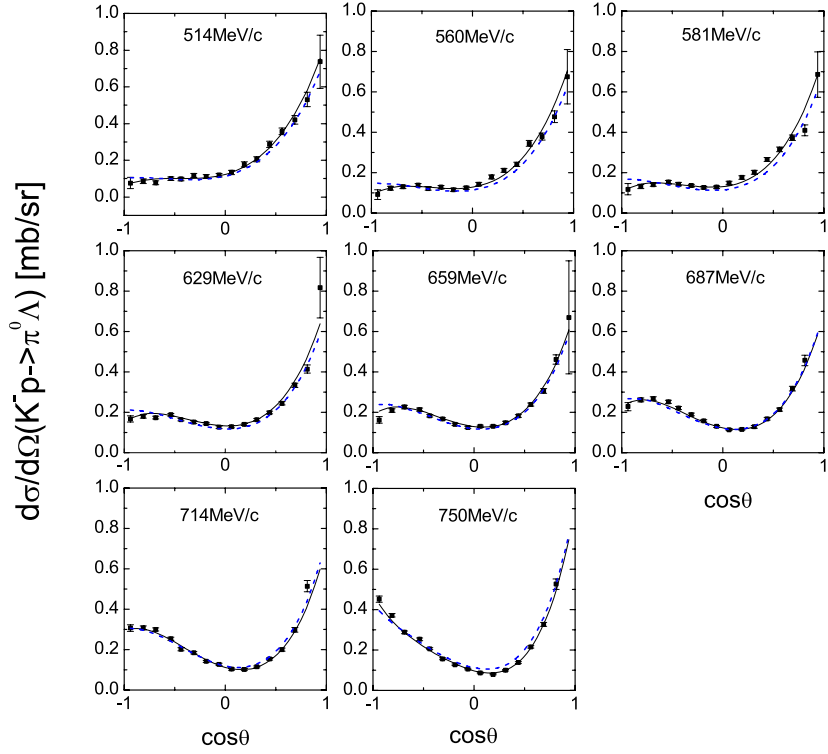}}
 \caption{The differential cross sections of the reaction $K^-+p\rightarrow
\pi^0+\Lambda$ compared with the experimental data~\cite{prakhov09},
where $\theta$ denotes the angle of the outgoing $\pi^0$ with
respect to beam direction in the c.m. frame. The dashed lines (blue)
are the best results with inclusion of only well established
(4-star) $\Sigma$ resonances in s-channel; the solid lines are best
results of including an additional $\Sigma({1\over 2}^+)$ resonance
in s-channel, with its mass near 1635~MeV and width around 121~MeV.
}
\end{figure*}

\begin{figure*}\label{fig:pol}
{\includegraphics[width=1.8\columnwidth]{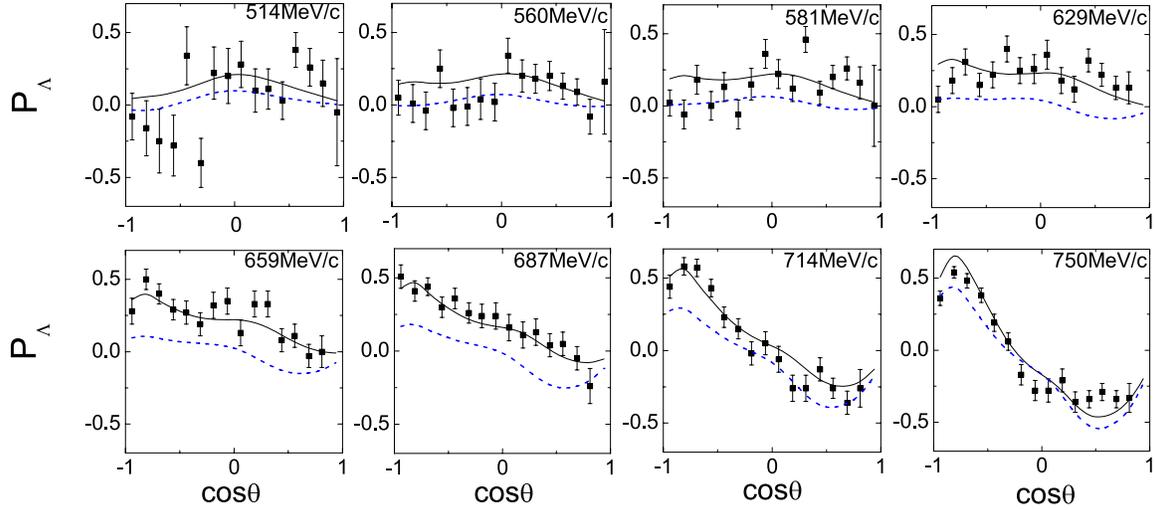}} \caption{Fits to
the $\Lambda$ polarization as a function of cos$\theta$ for the
reaction $K^-+p\rightarrow \pi^0+\Lambda$ compared with the
experimental data~\cite{prakhov09}, where $\theta$ denotes the angle
of the outgoing $\pi^0$ with respect to beam direction in the c.m.
frame. The dashed lines (blue) are results with inclusion of only
well established (4-star) $\Sigma$ resonances in s-channel; the
solid lines are results of including an additional $\Sigma({1\over
2}^+)$ resonance in s-channel with mass near 1635~MeV and width
around 121~MeV. }
\end{figure*}

For a better description of the data, we need to introduce some
other $\Sigma$ resonances in s-channel. We try them in the analysis
with their coupling constants, mass, and width as free parameters
and check if they are favored by the present data.

Among the $J^P={1\over 2}^\pm, {3\over2}^\pm$ $\Sigma$ resonances in
s-channel, our best fit comes from the inclusion of a $J^P={1\over
2}^+$ $\Sigma$ resonance with mass near 1635 MeV, and width around
121 MeV.

The solid lines in Fig.~2 and Fig.~3 shows this best fit compared
with the experimental data of the differential cross sections and
the $\Lambda$ polarization. The analysis includes 18 tunable
parameters in the allowed range and the $\chi^2$ for this best fit
is 223 for the total 248 data points. From the solid lines in Fig.~2
and Fig.~3, one can see that the experimental data can be much
better described with the inclusion of the $\Sigma({1\over 2}^+)$
resonance, especially with the $\Lambda$ polarization.

\begin{table*}
\caption{Adjusted parameters for high mass $\Sigma$ resonances,
which includes the $\Sigma(1635){1\over 2}^+$ resonance. Statistic
uncertainties and PDG estimates are also listed. }
\begin{tabular}{c|c|c|c}
\hline\hline   & mass(MeV)(PDG estimate) &
$\Gamma_\mathrm{tot}$(MeV)~(PDG estimate) &
$(\Gamma_{\pi\Lambda}\Gamma_{KN})^{1\over 2}/\Gamma_\mathrm{tot}$ ~(PDG range) \\
 \hline
$\Sigma(1670){3\over 2}^-$ & $1673.4^{+1.1}_{-0.8}$~(1665,1685)&
$54\pm 5$~(40,80)& $0.08^{+0.02}_{-0.015}~(0.02,0.17)$
\\\hline
$\Sigma(1635)$ or $\Sigma(1660){1\over 2}^+$ &
$1635^{+3}_{-4}(1630,1690)$ & $121^{+12}_{-10}(40,200)$&
$-0.064_{-0.015}^{+0.012}$(0,0.24) \\ \hline\hline
\end{tabular}
\end{table*}

\begin{table*}
\caption{Adjusted parameters with statistic uncertainties for the
couplings in t-channel, u-channel and s-channel $\Sigma(1189)$ and
$\Sigma^*(1385)$ exchange.} 
\begin{tabular}{c|c|c|c|c}
\hline\hline $g_{K^*N\Lambda}$~(model range) &
$g_{K^*N\Lambda}\kappa_{K^*N\Lambda}$~(model range)& $g_{\pi
NN}g_{KN\Lambda}$~(SU(3))             &
$g_{KN\Sigma}g_{\Sigma\Lambda\pi}$~(SU(3))    &
$f_{KN\Sigma^*}f_{\Sigma^*\Lambda\pi}$~(SU(3))  \\
 \hline
$-6.11^{+0.15}_{-0}$~($-6.11,-4.26$)~\cite{stoks99} &
$-11.33^{+0}_{-0.12}$~($-16.3,-10.4$)~\cite{stoks99}  &
$-177^{+9}_{-7}$~($-176$) & $49.1^{+0}_{-1}~(34.8)$ &
$-3.95^{+0.35}_{-0.38}$~($-4.1$)
\\ \hline\hline
\end{tabular}
\end{table*}

In Table I, we show the central values and statistic uncertainties
for 6 of the parameters on the s-channel $\Sigma$ resonances in this
energy range. From Table I, the mass of $\Sigma(1670)$ is precisely
around $1673$~MeV, and its width is around 54~MeV, which are
consistent with the PDG estimates~\cite{PDG08}. The coupling
constant lead $(\Gamma_{\pi\Lambda}\Gamma_{KN})^{1\over
2}/\Gamma_\mathrm{tot}\sim 0.08$ for $\Sigma(1670)$, which is
compatible with the PDG range. The mass of $\Sigma(1775)$ is much
larger than the c.m. energies of this experiment, and thus the fits
to the data are not sensitive to the parameters of $\Sigma(1775)$
within their PDG ranges. We fix the parameters of $\Sigma(1775)$ to
their PDG central values. The $J^P={1\over 2}^+$ $\Sigma(1635)$ from
our analysis is well in accordance with the 3-star
$\Sigma(1660){1\over 2}^+$ in PDG, giving a further support for the
existence of the resonance. Our fit with the $\Sigma(1635){1\over
2}^+$ is stable with the parameters well constrained. By including
this resonance, the $\chi^2$ drops from about 763 to 223 for the
total 248 data points.

The propagators of the s-channel $\Sigma$ resonance exchanges are
Breit-Wigner form in this analysis, where the mass and width of the
resonance are constant parameters. We also check the energy
dependent form of the Breit-Wigner propagator, \ie, to replace the
constant $i m_{\Sigma}\Gamma_{\Sigma}$ in the propagator by the
energy dependent form $i\sqrt{s}\Gamma(s)$. From the check on the
$\Sigma(1670){3\over 2}^-$, We find that the central values of the
mass and width of $\Sigma(1670)$ become larger by 2~MeV and 10~MeV,
respectively, while the effects on other parameters are very small.

The other 12 free parameters of this analysis includes 5 coupling
constants and 7 cutoff parameters in the form factors of the total 7
channels. In Table II we list the fitted results of the 5 free
parameters on the couplings of the t-channel, u-channel and
s-channel $\Sigma(1189)$ and $\Sigma(1385)$ contributions. Note that
the first two parameters are couplings of the t-channel $K^*$
exchange, and their ranges are constrained by the potential
model~\cite{stoks99}. We also check to widen the ranges of the two
parameters, and find only small shifts of the other parameters. Thus
the uncertainties on the parameters of the t-channel do not change
the main results of the analysis.

The research for the possible new $\Sigma({1\over 2}^-)$ near
1380~MeV has always been our concern, and previous work has shown
some evidence of it~\cite{zou08,wu09,gao10}. In this work, we also
check whether this data set is compatible with the existence of the
$\Sigma(1380)$. Without including the $\Sigma(1635)$, we try to
include a $\Sigma({1\over 2}^-)$, and constrain its mass above 1360
~MeV. From our analysis, the best fit gives $\chi^2=385$ a minimum
mass, a small coupling constant $g_{KN\Sigma({1\over
2}^-)}g_{\Sigma({1\over 2}^-)\pi\Lambda}\sim -1.26$ and width around
315~MeV. This shows that the existence of a $\Sigma({1\over 2}^-)$
near 1380~MeV with sizeable couplings is not ruled out by the
present data, although there is no strong evidence of it. This
result is understandable since 1380~MeV is much smaller than the
energy range of the experiment.

With the same procedure, we also try $\Sigma^*({3\over 2}^\pm)$
states in the analysis. The resulted $\chi^2$ are not significantly
improved as the case of the $\Sigma(1635){1\over2}^+$, which shows
no convincing evidence of their existences in the energy range of
the experiment.

Some uncertainty may still exist from the uncertainty in the
coupling constants and cutoffs, while the main results of this
analysis will not change.

As in previous analyses listed in PDG for the same reaction, here we
have not imposed the unitarity by a multichannel description. It is
clear that couple channel analysis is most appropriate way to
evaluate resonance properties. However we found that other relevant
channels are not well studied and there are no new precise data. In
principle that is a general problem. That is why in PDG the listed
analyses are also based on the single channel fit. Measurements with
wider energy ranges and combined channel analysis in the future will
be helpful to provide more information on properties and
interactions of the $\Sigma$ resonances.

\section{SUMMARY}

The neutral particles production from $K^-p$ interactions have been
measured by the Crystal Ball Collaboration for incident momentum of
$K^-$ between 514 and 750~MeV/$c$. Using the high precision new
data, we analyze the differential cross sections and the $\Lambda$
polarization of the reaction $K^-+p\rightarrow\pi^0+\Lambda$ with
the effective Lagrangian method. We include the contributions from
t-channel $K^*$ exchange, u-channel proton exchange, and four-star
$\Sigma$ resonances exchanges in s-channel, \ie, $\Sigma(1189)$,
$\Sigma(1385)$, $\Sigma(1670)$ and $\Sigma(1775)$ in our analysis.
We find that these 6 ingredients are still insufficient, with
$\chi^2\sim 763$ for the total 248 data points. We try to include
some new ingredient in our analysis and the best result is to
include a $\Sigma$ resonance with $J^P={1\over 2}^+$, mass near
1635~MeV and width around 121~MeV. The $\chi^2$ drops from 763 to
223 for the 248 data points. The properties of this resonance is
well in accordance with the 3-star $\Sigma(1660){1\over 2}^+$ in
PDG, providing a further support for the existence of the resonance.

\begin{acknowledgments}

This project is supported by the National Natural Science Foundation
of China under Grant 10905059, 10875133, 10821063, 11035006, the
Chinese Academy of Sciences under Project No.KJCX2-EW-N01, the China
Postdoctoral Science Foundation and the Ministry of Science and
Technology of China (2009CB825200).
\end{acknowledgments}

\end{document}